# Object-Oriented Intensional Programming: Intensional Classes Using Java and Lucid

Aihua Wu
Computer Science and Software Engineering
Faculty of Engineering and Computer Science
1455 de Maisonneuve Blvd. W.
SGW, EV8.245,
Concordia University, Montreal, Quebec,
Canada, H3G 1M8
aihua_wu@cse.concordia.ca

Joey Paquet
Computer Science and Software Engineering
Faculty of Engineering and Computer Science
1455 de Maisonneuve Blvd. W.
SGW, EV3.221,
Concordia University, Montreal, Quebec,
Canada, H3G 1M8
paquet@cse.concordia.ca

Serguei A. Mokhov
Computer Science and Software Engineering
Faculty of Engineering and Computer Science
1455 de Maisonneuve Blvd. W.
SGW, EV7.139-2,
Concordia University, Montreal, Quebec,
Canada, H3G 1M8
mokhov@cse.concordia.ca

**Abstract**

This article introduces Object-Oriented Intensional Programming (OO-IP), a new hybrid language between Object-Oriented and Intensional Programming Languages in the sense of the latest evolutions of Lucid. This new hybrid language combines the essential characteristics of Lucid and Java, and introduces the notion of object streams which makes it is possible that each element in a Lucid stream to be an object with embedded intensional properties. Interestingly, this hybrid language also brings to Java objects the power to explicitly express and manipulate the notion of context, creating the novel concept of *intensional object*, i.e. objects whose evaluation is context-dependent, which are here demonstrated to be translatable into standard objects. By this new approach, we extend the use and meaning of the notion of intensional objects and enrich the meaning of object streams in Lucid and semantics of intensional objects in Java.

## 1 Introduction

The origins of Intensional Programming (IP) is the Lucid programming language that dates back to 1974 [3]. After more than 30 years of development, history has proven that it is a programming paradigm whose languages are diversified and in constant evolution. However, Intensional Programming is an off-main-stream programming paradigm whose concrete applicability still needs to be proven in order to be widely accepted. On the other side, Object-Oriented languages are known by literally all computer scientists and software developers. Especially, Java is a very popular and widely used language in today's application domains. One way to increase the visibility of Intensional Programming is to make it more mainstream by a marriage between Object-Oriented Programming and Intensional Programming, allowing a broader audience to be exposed to the benefits of Intensional Programming.

The GLU (Granular Lucid) system, developed at the Stanford Research Institute in the 1990s, provided with a hybrid intensional-procedural by allowing Indexical Lucid programs to use C or Fortran data structures and functions in order to increase the granularity of data and operators manipulated by the Lucid part of its programs [7, 8]. It then relied on a Generator-Worker distributed execution architecture to execute such programs, exploiting the intrinsic parallel evaluation possibilities of Lucid made possible the purely functional nature of Lucid. Due to the lack of flexibility of its architecture, GLU was not able to adapt to further evolutions of Lucid, or to include interfacing with object-oriented languages [17].

The General Intensional Programming System (GIPSY) was designed using the same model as GLU, but with flexibility and adaptability in mind [18, 24, 20]. Using a framework approach, the GIPSY has been used to develop compilers for different variants of Lucid, which are allowed to coexist in the same program [27, 21, 12]. Moreover, its flexible design also permits Lucid programs to use procedures defined in virtually any procedural language. Although our current solution permits to have Lucid variants to use objects as first class values, i.e. a Lucid part that uses an object-oriented part, we do not yet have the possibility to have the object-oriented part to use Lucid constructs. Providing both of these possibilities would bring forth the first "Object-Oriented Intensional Programming Language".

**Problem Statement** The scientific contribution of this paper aims at integrating the Object-Oriented and Intensional Programming paradigms into a hybrid syntax and computational model executable in a distributed environment, allowing both (1) the use of objects as first class values in Lucid and (2) the use of Lucid inside of classes, thus defining *intensional classes*.

**Proposed Solution** In order to provide a fully integrated OO-IP hybrid language, we need to provide integration of objects into Lucid, as well as Lucid into objects. Our solution aims at: (1) allowing Lucid to define streams of objects, and to introduce the dot notation in Lucid syntax, allowing Lucid to use objects and their members; (2) allowing classes to define *intensional data members*, as well as allowing methods to use intensional expressions as part of their statements, yielding *intensional methods*; implementation of the proposed solution inside the GIPSY infrastructure, thus permitting (3) the introduction of any flavor of Lucid inside of class declarations by the automated translation of Lucid variants into Generic Lucid prior to execution, as well as (4) the execution of such hybrid OO-IP programs in a scalable distributed environment using the GIPSY's run-time engine architecture. The big picture of our particular vision of OO-IP can be itemized as the following:

- The object-oriented "dot" notation is introduced in Lucid to access data and function elements of a class similarly as it was first introduced in [12, 11].
- Intensionality is introduced into Java classes by the embedding of Lucid expressions into otherwise standard Java classes.
- Java classes encapsulate the notations of intensionality and dimensionality.
- For an intensional class varying over a multidimensional manifold, one instance of this class exists for each point in this manifold.
- As intensional members are declared using Lucid expressions and translated into standard Java classes calling the eduction engine for the evaluation of their embedded Lucid expressions, the standard syntax and semantics of Java is preserved.
- An intensional data member in fact declares an intensional relationship applying to the stream of values that it declares across the objects of the class.
- All intensional members' object instances are dynamically generated with the invocation of the intensional execution engine, driven by the current context and the Lucid expression defining that intensional member.

- All Java classes including Lucid expressions imply intensional evaluation only when evaluating the part of their definition that embeds Lucid expressions. Standard Java classes are still evaluated using the standard Java semantics, i.e. the execution of OO-IP programs is driven by standard Java semantics, unless in the punctual presence of Lucid expressions, in which cases the evaluation is switched to the intensional evaluation engine.
- The hybrid language being proposed here is bound to the Java syntax and semantics, so we thus name it "JOOIP".

# 2 Syntax and Semantics of JOOIP

We describe here the syntax of JOOIP, explaining (1) how can Lucid use objects' members, and (2) how can Lucid expressions be embedded into Java classes. We also give a set of operational semantic rules corresponding to (1), extending the standard Lucid operational semantics as defined in [17], as well as translation rules for the translation of JOOIP code into standard Java code.

## 2.1 Syntax

In order to achieve the stated features set in Section 1, we have determined that the solution is composed of two separate and complementary aspects. One is about the change to the original syntax of Lucid in order to allow Lucid syntax to manipulate objects' members using the standard "dot notation" used in object-oriented languages. That has already been achieved in various solutions. The other is about the change to the original syntax of Java to allow Java classes to define *intensional data members*, as well as *intensional methods*. Syntactically, both are achieved by allowing the embedding of Lucid expressions inside of an otherwise standard Java class. In Table 1, productions (13)-(24) provide the first part, and productions (1)-(12) shows the second part.

In Table 1, productions (19) and (20) are added into the original Lucid syntax to allow Lucid to access, respectively, data and function elements of a class as it was done in Objective Lucid in [12, 11]. The syntax of JOOIP shows in productions (1)-(12) how to integrate the Lucid expressions into a Java class. We use a tag "`/@`" to start an embedded Lucid expression, which is to be ended by a corresponding "`@/`" tag (see production (5)). Preceded by "`#`", the <Lucid variant tag> indicates which Lucid variant is used for the following embedded Lucid expression (see production (6)-(12)). This will enable our preprocessor to send the Lucid code segment to the corresponding Lucid compiler during translation time. These are the same tags and design solution used by the `Preprocessor` of `GIPC` in generalized compilation [10]. This solution permits the JOOIP to embed Lucid expressions written in any variant of Lucid supported by the GIPSY, a unique feature of our solution.

## 2.2 Semantics

According to [17], intensional programs are evaluated in two environments:

- a definition environment $\mathscr{D}$ that is constructed as the program is compiled. The definition environment is an abstract equivalent to compilers' symbol table. It is a partial function that maps identifiers to their definitions.
- a context environment $\mathscr{P}$ defining at what point in the multidimensional context space an expression is to be evaluated. It is a set of mappings, associating each dimension with an index over this dimension. Together, the mappings define a point in the multidimensional context space defined by the program.

Table 1: Syntax of JOOIP

| | | | |
|---|---|---|---|
| | | | *... standard Java syntax* |
| (1) | <data member declaration> | ::= | *(standard Java data member declaration)* |
| (2) | | \| | type id = <embedded Lucid expr>; |
| (3) | <expression term> | ::= | *(standard Java expression terms)* |
| (4) | | \| | <embedded Lucid expr> |
| | | | *standard Java syntax...* |
| | | | |
| (5) | <embedded Lucid expr> | ::= | /@<Lucid variant tag> <E> @/ |
| (6) | <Lucid variant tag> | ::= | # <Lucid variant id> |
| (7) | | \| | $\varepsilon$ |
| (8) | <Lucid variant id> | ::= | GIPL |
| (9) | | \| | INDEXICALLUCID |
| (10) | | \| | JLUCID |
| (11) | | \| | OBJECTIVELUCID |
| (12) | | \| | LUCX |
| | | | |
| (13) | <E> | ::= | id |
| (14) | | \| | <E>(<E>,...,<E>) |
| (15) | | \| | if <E> then <E> else <E> |
| (16) | | \| | #<E> |
| (17) | | \| | <E>@[<E>:<E>,...,<E>:<E>] |
| (18) | | \| | <E> where <Q> end |
| (19) | | \| | <E>.id |
| (20) | | \| | <E>.id(<E>,...,<E>) |
| (21) | <Q> | ::= | dimension id,...,id |
| (22) | | \| | id=<E> |
| (23) | | \| | id(id$_1$,...,id$_n$)=<E> |
| (24) | | \| | <Q><Q> |

The standard operational semantic rules of generic Lucid and JLucid from [17, 12, 6] are extended as shown in Figure 1. We present here only the rules that require changes from the standard semantic rules, as well as additional rules related to features uncovered by generic Lucid. In these semantic rules, the semantic operator † represents the addition of a mapping in the definition environment $\mathscr{D}$, associating an identifier with its corresponding semantic record, here represented as a tuple. Following is detailed textual description of the meaning of each semantic rule:

**J$_{\textbf{CDef}}$** semantically identifies a Java class by the syntactical form: class cid{...}, associates this class declaration to the identifier cid, and stores it in the definition environment $\mathscr{D}$ as the semantic record (class, cid, JavaCDef). A class can contain member variables (JavaVDef) and member functions (JavaFDef, also called "methods"). These are processed in a similar manner by the two following semantic rules.

**J$_{\textbf{VDef}}$** semantically identifies a Java class member variable (or data member) in a Java class JavaCDef by the syntactical form: public type vid... found inside of this class declaration. The semantic record (classV, cid.vid, JavaVDef) is used to represent a Java class data member vid declared inside a class declaration JavaVDef for the class cid.

$$\mathbf{L_{objV}} : \frac{\begin{array}{l}\mathscr{D},\mathscr{P} \vdash E : v \quad \mathscr{T}(v) = \mathscr{D}(cid) = (\texttt{class, cid, } \underline{\texttt{JavaCDef}}) \\ \mathscr{D},\mathscr{P} \vdash vid : vid \quad \mathscr{D}(cid.vid) = (\texttt{classV, cid.vid, } \underline{\texttt{JavaVDef}}) \\ \mathscr{D},\mathscr{P} \vdash \mathbf{JVM}[[v.vid]] : v_r\end{array}}{\mathscr{D},\mathscr{P} \vdash E.vid : v_r}$$

$$\mathbf{L_{objF}} : \frac{\begin{array}{l}\mathscr{D},\mathscr{P} \vdash E : v \quad \mathscr{T}(v) = \mathscr{D}(cid) = (\texttt{class, cid, } \underline{\texttt{JavaCDef}}) \\ \mathscr{D},\mathscr{P} \vdash fid : fid \quad \mathscr{D}(cid.fid) = (\texttt{classF, cid.fid, } \underline{\texttt{JavaFDef}}) \\ \mathscr{D},\mathscr{P} \vdash E_1,\ldots,E_n : v_1,\ldots,v_n \\ \mathscr{D},\mathscr{P} \vdash \mathbf{JVM}[[v.fid(v_1,\ldots,v_n)]] : v_r\end{array}}{\mathscr{D},\mathscr{P} \vdash E.fid(E_1,\ldots,E_n) : v_r}$$

$$\mathbf{L_{FF}} : \frac{\begin{array}{l}\mathscr{D}(ffid) = (\texttt{freefun, ffid, } \underline{\texttt{JavaFFDef}}) \\ \mathscr{D},\mathscr{P} \vdash E_1,\ldots,E_n : v_1,\ldots,v_n \\ \mathscr{D},\mathscr{P} \vdash \mathbf{JVM}[[ffw.ffid(v_1,\ldots,v_n)]] : v_r\end{array}}{\mathscr{D},\mathscr{P} \vdash ffid(E_1,\ldots,E_n) : v_r}$$

$$\mathbf{J_{CDef}} : \frac{\underline{\texttt{JavaCDef}} = \texttt{class cid } \{\ldots\}}{\mathscr{D},\mathscr{P} \vdash \underline{\texttt{JavaCDef}} : \mathscr{D}\dagger[\texttt{cid} \mapsto (\texttt{class, cid, } \underline{\texttt{JavaCDef}})],\ \mathscr{P}}$$

$$\mathbf{J_{VDef}} : \frac{\begin{array}{c}\underline{\texttt{JavaCDef}} = \texttt{class cid } \{\ldots \underline{\texttt{JavaVDef}} \ldots\} \\ \underline{\texttt{JavaVDef}} = \texttt{public type vid } \ldots;\end{array}}{\mathscr{D},\mathscr{P} \vdash \underline{\texttt{JavaVDef}} : \mathscr{D}\dagger[\texttt{cid.vid} \mapsto (\texttt{classV, cid.vid, } \underline{\texttt{JavaVDef}})],\ \mathscr{P}}$$

$$\mathbf{J_{FDef}} : \frac{\begin{array}{c}\underline{\texttt{JavaCDef}} = \texttt{class cid } \{\ldots \underline{\texttt{JavaFDef}} \ldots\} \\ \underline{\texttt{JavaFDef}} = \texttt{public ft fid}(\texttt{fpt}_1\ \texttt{fp}_1,\ \ldots,\ \texttt{fpt}_n\ \texttt{fp}_n)\{\ldots\}\end{array}}{\mathscr{D},\mathscr{P} \vdash \underline{\texttt{JavaFDef}} : \mathscr{D}\dagger[\texttt{cid.fid} \mapsto (\texttt{classF, cid.fid, } \underline{\texttt{JavaFDef}})],\ \mathscr{P}}$$

$$\mathbf{J_{FFDef}} : \frac{\begin{array}{c}\underline{\texttt{JavaFFWCDef}} = \texttt{class ffw } \{\ldots \underline{\texttt{JavaFFDef}} \ldots\} \\ \underline{\texttt{JavaFFDef}} = \texttt{ft ffid}(\texttt{fpt}_1\ \texttt{fp}_1,\ \ldots,\ \texttt{fpt}_n\ \texttt{fp}_n)\{\ldots\}\end{array}}{\mathscr{D},\mathscr{P} \vdash \underline{\texttt{JavaFFDef}} : \mathscr{D}\dagger[\texttt{ffid} \mapsto (\texttt{freefun, ffid, } \underline{\texttt{JavaFFDef}})],\ \mathscr{P}}$$

Figure 1: Additional basic semantics of JOOIP

$\mathbf{J_{FDef}}$ semantically identifies a Java member function in a Java class $\underline{\texttt{JavaCDef}}$ by the syntactical form: $\texttt{public ft fid}(\texttt{fpt}_1\ \texttt{fp}_1,\ \ldots,\ \texttt{fpt}_n\ \texttt{fp}_n)\{\ldots\}$.
The semantic record $(\texttt{classF}, \texttt{cid.fid}, \underline{\texttt{JavaFDef}})$ is used to represent a Java member function `fid` declared inside a class declaration $\underline{\texttt{JavaCDef}}$ for the class `cid`.

$\mathbf{J_{FFDef}}$ semantically identifies a "Java free function" (i.e. a free function such as in C++, but written in Java syntax) that is not explicitly defined in a given class, and has the syntactical form:
$\texttt{ft ffid}(\texttt{fpt}_1\ \texttt{fp}_1,\ \ldots,\ \texttt{fpt}_n\ \texttt{fp}_n)\{\ldots\}$. The semantic record $(\texttt{freefun}, \texttt{ffid}, \underline{\texttt{JavaFreeFDef}})$ is used to represent a "Java free function" `ffid`, i.e. a function that is directly available in the Lucid program, and that is not a member of any class. Note that free functions are not allowed in standard Java. In terms of implementation, these "free functions" are all put inside a wrapper class to be part of the worker component of the execution engine as originally defined in [12, 6].

$\mathbf{L_{objV}}$ defines the semantics of the evaluation of a reference to a class data member by a Lucid expression using the object-oriented dot operator. The top part of the rule insures that, in $E.vid$: (1) the Lucid expression $E$ evaluates to a value $v$ that is an object of type `cid`, as being associated in the definition environment $\mathscr{D}$ to the tuple $(\texttt{class}, \texttt{cid}, \underline{\texttt{JavaCDef}})$, (2) the variable `vid` is a public member of the class `cid`. Once this is established as holding, the Java Virtual Machine can be called upon to evaluate $v.vid$ (noted as $\mathbf{JVM}[[v.vid]]$), to yield a value $v_r$.

```
(TR1) Start =>
      (1)  new file: cid.java
      (2)  new StringBuffer: HdBuf, IdentifierBuf, StaticBuf, MethodBuf, JavaBuf
      (3)  HdBuf          << "import gipsy.lang.*;
                              import gipsy.GEE.GEE;
                              public class cid implements ISequentialThread{"
      (4)  IdentifierBuf  << "private GIPSYContext oContext;"
      (5)  StaticBuf      << "static {"
      (6)  MethodBuf      << "public WorkResult work(){cid.main(null); return null;}"

(TR2) standard Java syntax  =>
      (7)  JavaBuf        << standard Java data member declaration
      (8)  JavaBuf        << standard Java expression terms

(TR3) <data member declaration> ::= qualifier type id = <embedded Lucid expr>   =>
      (9)  IdentifierBuf  << "private type id =0;
                          <<  private Boolean b(id)IsWritten = false;"
      (10) IdentifierBuf  <<  see (TR4);

(TR4) <embedded Lucid expr> ::=  @/<Lucid variant tag> <E>/@  =>
      (11) IdentifierBuf  << "private static GIPSYProgram SoGEERi;
                          <<  public GEE oGEEi = new GEE(SoGEERi);"
      (12) StaticBuf      << "GIPC oGIPCi = new GIPC(<E>);
                          <<  oGIPCi.compile();
                          <<  SoGEERi = oGIPCi.getGEER();"
      (13) JavaBuf        << "oGEEi.eval(oContext);"

(TR5) Finish   =>
      (14) StaticBuf      << "}"
      (15) cid.java       << HdBuf
      (16) cid.java       << IdentifierBuf
      (17) cid.java       << StaticBuf
      (18) cid.java       << MethodBuf
      (19) cid.java       << JavaBuf
      (20) cid.java       << "}"
```

Figure 2: Translation rules to translate JOOIP into Java

$\mathbf{L_{objF}}$ defines the semantics of the evaluation of a reference to a class member function by a Lucid expression using the object-oriented dot operator. The top part of the rule insures that, in $E.fid(E_1,\ldots,E_n)$: (1) the Lucid expression $E$ evaluates to a value $v$ that is an object of type `cid`, as being associated in the definition environment $\mathscr{D}$ to the tuple $(\texttt{class}, \texttt{cid}, \underline{\texttt{JavaCDef}})$, (2) the method `fid` is a public member of the class `cid`. Once this is established as holding, all actual parameters are evaluated to values $v_1,\ldots,v_n$, the JVM can be called upon to evaluate $v.fid(v_1,\ldots,v_n)$ (noted as $\mathbf{JVM}[[v.fid(v_1,\ldots,v_n)]]$), to yield a value $v_r$.

$\mathbf{L_{FF}}$ defines the semantics of the evaluation of "Java free functions". The rule is a simpler version of $\mathbf{L_{objF}}$ with no class type identifiers present, and no object to compute upon which to call the function. As mentioned earlier, Java does not have free functions, so we cannot claim that the Java Virtual Machine can execute them. In fact, all free functions are wrapped in a "free function wrapper" class at compilation, with all free functions inserted in it as static functions [12, 6]. The $\mathbf{J_{FFDef}}$ rule is inserting all the free functions in this wrapper class, which we called `ffw`. Then, upon calling such "free functions", this rule is called and assumes that the "free functions" have been wrapped as static functions into the `ffw` class, then call the appropriate function. This mechanism is an improvement and refinement over [12, 6].
Figure 2 shows the translation rules applied to translate JOOIP code into standard Java code. The following paragraphs explain each of the translation rules.

`TR1` When starting to translate a JOOIP program, the translator first creates a new Java file and four string buffers, the latter being rendered necessary by the fact that the translation output cannot be done sequentially upon sequential scanning of the input JOOIP program. The first translation step is to generate the class declaration header part to the Java file `(3)`, then generate the code for the instance *context* member of the intensional class `(4)`, then initiate a static block that is to contain static initialization of one `GEER` data member per each Lucid expression embedded in the intensional class `(5)`. Finally, a `work()` method is generated that is a simple wrapper that calls the `main()` function of the intensional class; this method is the one called by the GIPSY run-time system when run from within a GIPSY instance as prescribed by the `ISequentialThread` interface that the class is declared to implement at `(3)`.

`TR2` This rule signifies that the translator simply copies any standard Java code (i.e. code for which no translation rule applies) as-is to `JavaBuf`.

`TR3` This rule signifies the translation to occur when an intensional data member is encountered by the translator. On the surface, the processor will replace the intensional data member declaration with a standard Java data member declaration. For each such occurrence, the processor records the identifier in a table and leave a field for later link to the corresponding GEER as processed by `(TR4)`. As translation rule `(TR5)` takes effect, it will replace the Lucid expression by a call to the GEE to execute the Lucid expression. The `qualifier` is one of the Java's visibility qualifiers, such as `private`, `protected`, or `public`, where we currently opting out for `private`. The `type` denotes a Java member type that Java part expects the Lucid expression to produce. At run-time, when the Lucid expression is done evaluating, the Lucid-to-Java type matching and conversion occurs according to the definition of the type system presented in Section 3.2.

`TR4` This rule signifies the translation of the embedded Lucid code segments, as identified by the occurrence of the opening `@/` and closing `/@` markers. For each such occurrence, the translator generates an identifier of type `GIPSYprogram` (representing a GEER) and generates code in the `StaticBuf` to generate the instance GEER at run time by calling the appropriate parser in the GIPC framework, as directed by the Lucid variant tag. The translator then effectively replaces the embedded Lucid expression invocation by a call to the GEE that evaluates this Lucid expression in the context of the object in question.

`TR5` Finally, the translator aggregates all generated parts in the proper order to finish the generation of the intensional class in pure Java.

# 3 Implementation Details of the JOOIP Compiler

The JOOIP hybrid language is designed as a 2-way hybrid: we allow the embedding of Lucid code in the Java language and the Lucid code may refer to Java objects' members. Thus, at the implementation level, we design the modules in such a way as to make sure that the JVM can work smoothly with the GIPSY run-time engine. To avoid changing the Java compiler, we have chosen to implement the translation process as a multi-pass compiler that translates the JOOIP classes into pure Java classes, according to the semantics and translation rules presented earlier in Section 2. This compiler is fully integrated into the GIPSY, and, thus, is a GIPSY compiler instance in the GICF framework. Integrating our solution into the GIPSY permits us to reuse the existing Lucid compiler components, as well as executing the Lucid parts using the GIPSY's run-time environment, the GEE. The compilation process can be summarized as the following:

1. Firstly, the translator, as it parses a JOOIP program, identifies all embedded Lucid expressions and conceptually replaces them by a regular Java variable `Lucid_expr_i`, where `i` means it is the `i`-th embedded Lucid expression placed in the JOOIP. Concretely, in order to yield a value from the evaluation of this Lucid expression, the occurrence of the Lucid expression is replaced by a call to the GEE for the evaluation of this compiled Lucid expression (which has a corresponding GEER) by the GEE in the context `cxt` associated with the object in question. At the same time, the translator creates entries in a Lucid identifier symbol table to record `Lucid_expr_i`, as well as its corresponding Lucid expression in a form of a pointer to the root node of that expression in the AST. After all such embedded Lucid expressions are translated in this manner, the JOOIP program becomes an intermediate Java program, which can be syntactically parsed by a standard Java parser, but the parser will not be able to capture the special meaning for the `Lucid_expr_i` variables.

2. Then, the translator passes the intermediate Java program generated in the first step to our customized Java parser generated from the Java 1.5 grammar specification (found in [26]) altered to allow extraction of the Lucid code segments in order to get the symbol table that records Java class data members and becomes aware of the `Lucid_expr_i` corresponding to an *intensional data member declaration* or an *intensional expression term*. After learning that some `Lucid_expr_i` corresponds to an intensional data member declaration, the processor will apply translation rule `(TR3)` to generate standard Java code. If the `Lucid_expr_i` corresponds to an intensional expression term, the processor will apply translation rule `(TR4)` to generate standard Java code.

3. Next, the translator passes each embedded Lucid expression to the corresponding Lucid compiler, whose parser name is specified by the *Lucid variant id* tag as signified in translation rule `(TR4)`. The tag identifiers here correspond exactly to the language ID tags as defined in GICF [12] of the language parsers currently available in the GIPSY. In GICF *Lucid variant id* is known as LANGID and is presented in the syntax rules (8)-(12) in Table 1. Every Lucid compiler will return a GEER for each `Lucid_expr_i`, which is stored in the Lucid identifier table.

4. After that, the translator does rudimentary semantic checking only for the embedded Lucid expressions (the Java complete code checking is deferred to later analysis by the standard Java compiler). The limited semantic checking at this stage involves checking the symbol tables produced in the steps 1 and 2 for undefined or multiply defined identifiers in their corresponding scope according to the Lucid and Java semantics. This is especially important when a plain Java identifier is used in an intensional expression.

5. Finally, the translator generates a pure Java program, in which the `work()` method automatically generated is the one to be called by the GIPSY run-time system, thereby correctly linking this generated class with other such classes through the run-time system. It is also possible to invoke the pure Java program by itself as it defines the `main()` method to be the starting point of the computation, that eventually may invoke the GEE if the compiled Lucid expression is encountered at run-time. In such a case, the invocation of the GIPSY run-time (GEE) is deferred until some later point when an `eval()` call is made to the GEE. For it it work, the GIPSY compiled code should be in the CLASSPATH of our generated program at this step to be runnable.

To explain the implementation details of the compilation process described above, we show a class diagram of the JOOIP compiler in Figure 3. The `main()` method starts in the `JOOIPCompiler` class, and then, by using available Lucid parsers in the GIPSY, `JavaParser` and `SemanticAnalyzer`, which are already in the `GIPC` framework, it produces a compiled standard Java class. On the way, we can get all information in the symbol hashtable and ready for the `GEE` to use when we call it indirectly through the intensional code segments. Just like other compilers in the framework, the large portion of the JOOIP compiler was generated using JavaCC [26] and the Java 1.5 grammar specification written in `java15.jjt`. The grammar file was altered to allow for the intensional code segments to be extracted as well as the identifiers as described in the steps above. Later the compiler was customized with some GIPSY-specific annotations to allow it to be a part of the GIPC framework.

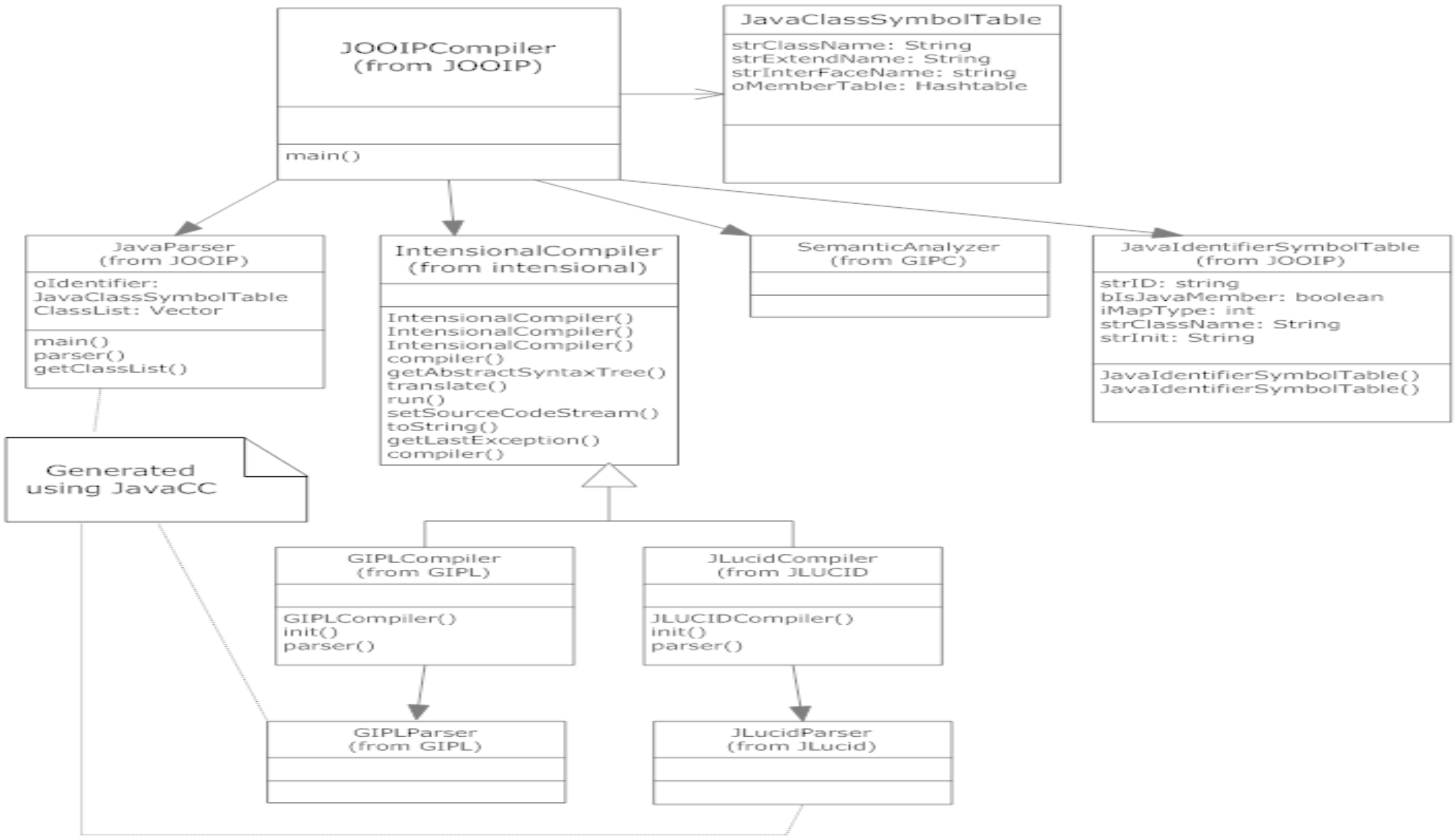

Figure 3: Class diagram of the JOOIP processor

## 3.1 Data Structures

The main data structure used is a symbol table which records all Java classes, variables, and Lucid identifiers tentatively called `JavaClassSymbolTable`. The detailed design of this table is shown in Figure 3. The string `strClassName` records the fully-qualified Java class name. The string `strExtendName` records the parent's class name. The string `strInterFacename` records the interface names. The hashtable `oMemberTable` records all members of this Java class. The detailed design of each identifier instance in the `oMemberTable` referred to a type class `JavaIdentifierSymbolTable`, is also presented in Figure 3. While designing our JOOIP language and early testing its properties we opted to have these two classes implemented ourselves. In `JavaIdentifierSymbolTable`, the string `strID` records the variable's name. The Boolean `bIsJavaMember` records if this variable is a plain Java class member or is an intensional identifier, which is defined in the Java class in the form of a Lucid expression. The integer `iMapType` records the data type of this variable, which is used in type matching between Lucid expression and Java types. Further details about the type mapping can be found in Section 3.2. The string `strClassName` records the fully-qualified Java class name this variable or intensional identifier belongs to. It is added here to simplify the semantic scope check for the Lucid code segments. The string `strClassInit` records the declaration of this variable, if it is an intensional identifier, the `strClassInit` will represent the intensional source code fragment. The `oEntry` member[1] records the entry of an AST of the intensional identifier. This AST will be ready only after the Lucid code segments are passed to the corresponding Lucid parsers and collected back as ASTs for re-linking. The dictionary `oLucidIdentifierDictionary`[2] records the semantic table for each intensional identifier. The final preparation of the dictionary instance completes only after the semantic checking of Lucid code is performed. In this hashtable, the `oEntry` and `oLucidIdentifierDictionary` will have value only when the boolean `bIsJavaMember` is set to `false`.

### 3.2 Type System

Just like the predecessor hybrid intensional-imperative languages studied in the GIPSY environment, JLucid or Objective Lucid, the JOOIP language cannot avoid the type mapping between Java and Lucid as the data type sets of both languages are not identical. In Table 2 we summarize the type mapping between Java and GIPSY types and the intensional counterpart. A particularity here, that any intensional dialect has a `dimension` type and its implementing class `GIPSYContext` in the GIPSY type system, but its value cannot be directly mapped to a specific Java type in JOOIP, and the dimensions can presently be, but may not be limited to in the future, integers and strings and have the corresponding types of tag sets [19, 25] attached. This especially concerns the Java method parameters and return types. Additionally, each intensional class instance has its own context managed internally as described earlier in the `cxt` data member of the generated Java class. Instead of doing any re-mapping to the Java type in that regard, we simply kept the object context reference as `GIPSYContext`. The type matching and conversion happens at run-time, when the GEE completes the evaluation of an intensional expression and comes back with a value. The final GIPSY type of the value is determined, examined, and, if possible, converted by the Lucid-Java adapter just before returning to the calling Java class. A run-time type check semantic error may happen at this point in type if it is not possible to match the type according to the table shown. A similar process is invoked in the other direction of evaluation, when the intensional segment uses the result of an imperative evaluation, which was broadly discussed in JLucid and Objective Lucid works earlier. For an in-depth discussion of the GIPSY type system, please refer to [13].

## 4 Discussions on JOOIP

JOOIP combines Lucid and Java in a different (inverse) way than was done in [12, 11, 6]. We can say it is Lucid conservatively extended with intensional objects, class hierarchy with inheritance, and encapsulation; it is an object-oriented imperative language, enhanced with explicit context-aware members. In this section, we discuss JOOIP from both object-oriented aspect and intensional aspect.

[1]The `SimpleNode` class is a customized version of the a class generated by any JavaCC-generated [26] parsers with ASTs.

[2]the `Dictionary` class is developed in-house of the GIPSY project and is in use by the `GIPC` and `GEE` components and their subcomponents.

Table 2: Summary of type mappings between Java and Lucid in JOOIP

| Lucid type | Java type | GIPSY Type System |
|---|---|---|
| dimension | int, String | `GIPSYContext` |
| char | char | `GIPSYCharacter` |
| int | byte | `GIPSYInteger` |
| int | short | `GIPSYInteger` |
| int | int | `GIPSYInteger` |
| (–) | long | `GIPSYInteger` |
| float | float | `GIPSYFloat` |
| double | double | `GIPSYDouble` |
| bool | boolean | `GIPSYBoolean` |
| [] | array | `GIPSYArray` |
| string | String | `GIPSYString` |
| object | class | `GIPSYObject` |
| (–) | interface | (–) |
| (–) | enum | `GIPSYObject` |
| bool:true | void | `GIPSYVoid` |

## 4.1 OOIP vs. IOOP

There are two ways to introduce intensional and dimensional concepts into objects. One is to devise new Lucid constructs to allow the declaration of "Lucid classes"; another solution would be to allow the use of Lucid syntactical constructs into classes (e.g. Java classes). In the following sections, we briefly introduce these two concepts and come up with arguments enabling us to make a choice between the two approaches.

### Classes in Lucid

This approach would require changing the syntax of Lucid to allow class declarations to be expressed using the Lucid syntax. This might seem an interesting concept, and we have been lured into it in the course of developing JOOIP. Eventually, however, we have figured out different important reasons why this is not a viable solution. The main reason is that Lucid is a typeless language, i.e. it does not declare types explicitly, nor does it explicitly refer to types anywhere in its syntax. Lucid's semantics and execution model use type inference and type analysis, but types are not explicitly stated in Lucid programs. As classes are abstract data types, introducing classes inside of Lucid would introduce explicit reference to types into Lucid, and we felt that this would considerably change the language, up to a point where the new language would cease to be in the "standard" Lucid family of languages, and would thus require that our existing execution engine and compiler framework to be considerably redesigned and have their implementation changed. For these important reasons, this solution was never attempted, and we also chose to reject it.

### Lucid in Classes

This approach requires an extension of the syntax of Java so that Lucid code is allowed to be included inside Java classes, creating *intensional classes*. Here the difficulty lies not so in the syntax, but rather in the semantics of the thus created language. Java has a very well established and standard semantics as embedded in Java Virtual Machines. We certainly do not want to have to dig deep and bury ourselves in Java semantics and JVM implementation. A workaround solution to this problem comes in providing a translation of the embedded Lucid constructs into standard Java, thus having the intensionality syntactically expressed in Lucid whose semantics rules are semantic translation rules that translate such constructs into standard Java constructs, thus allowing ourselves to use the underlying Java semantics rather than change it. In the following sections we present some interesting concepts of importance related to the integration of the notion of object along with the notion of intensionality.

## 4.2 Object Mutability

Object mutability, i.e. the fact that an object's state or behavior can vary in time, is of central importance in the design of a hybrid OO-IP language. We discuss here the notions of mutable/immutable objects, as well as the notion of *context-mutability* i.e. the fact that, given that our OO-IP objects are defined in a multidimensional context space, some of these objects are to be constant or mutable across the context space i.e. *context-immutable* or *context-mutable*.

### Mutable objects

"Mutable object" means when one has a reference to an instance of an object, the contents or state of that instance (i.e. the values of its data members) can be altered, thus making it so that the same object changes its state in time [23], and even that its methods would expose an apparently non-functional behavior over time, if the methods are referring to the changing values of its data members. The use of mutable objects with Lucid is highly problematic because the semantics of Lucid assumes that the values associated with expressions are constant in a given context, and thus only have to be evaluated once, thus permitting the storage of evaluation results, and their retrieval when the same expressions are to be evaluated. Permitting to have streams of mutable objects would thus be semantically invalid, as each stream element's value could possibly change in time. One interpretation to mutable objects from a valid IP perspective would be that mutable objects could be represented as a stream of immutable objects, where the stream is "recording" the changes in an object's state in the *time dimension*. That, in fact, brings us to the absolute beginning of Lucid, where Lucid streams were used to represent the changes of values of program variables upon and throughout execution, in the context of program verification [1, 2].

### Immutable objects

Immutable objects are simply objects whose state (the object's data members values) does not change after construction [14]. Immutable objects thus expose strict functional behavior throughout their entire lifetime. An immutable object is an object whose evaluation of any of its member always return the same value, behaving like as if all its members would be preceded by `const` in C++ or `static final` in Java, and all its functions always return the same value when given the same parameters. Immutable objects work nicely in our view of OO-IP, because of the converse of the reasons provided in the previous section discussing mutable objects.

### Context-immutable objects

This is a concept arising from the field we are discussing in this research, i.e. objects whose evaluation is context-independent. In Intensional Programming, expressions are evaluated in a multidimensional context, possibly yielding a different value for the same expression when evaluated in different contexts. In Object-Oriented Programming, expressions evaluate to objects, and there is no such thing as the context of evaluation. Combining these two paradigms, a context-immutable object is an object whose evaluation is independent on the context of evaluation, i.e. an object that is invariant in a multidimensional context space. This means that a "context-immutable object" does not embed any intensional expression, or it may be an object that embeds an intensional expression, but that expression is context-invariant. We can say any traditional Java object is a context-immutable object.

**Context-mutable objects**

An object of this classification is one whose evaluation is dependent on the context of evaluation. This means a context-mutable object has an embedded dimensional concept in its class declaration and that the evaluation of its members varies in a multidimensional context space, i.e. this class embeds at least one Lucid expression. In our approach, intensional objects are declared by the inclusion of Lucid expressions in their class declarations, either by having data members declared as a Lucid expression, or having member functions to embed Lucid expressions as part of their Java statements and expressions. By this way, we can permit both mutable and immutable objects by embedding or not embedding such Lucid expressions in their class declarations. Interestingly, as soon as the context of evaluation is decided, objects are evaluated to context-immutable objects which are normal Java objects.

## 4.3 Demand-Driven Constructors

Object-oriented programming uses constructors to generate object. However, Lucid uses demand-driven evaluation, and may generate a demand for objects that has not been constructed yet. We cannot design constructors for Lucid streams because most Lucid streams are infinite. However, we translate *intensional data members* in an *intensional class* to standard Java data members by adding a constructor that calls the eduction engine for the creation of the new object. When there is a demand for an object, the constructor of the corresponding class is called, potentially generating demands for the computation of each *intensional attribute* in this class. For example, if each element on the stream *S* is an object. Then "`S@[d:3]`" generates a demand for an object, which generates other demands on evaluation at context: `[d:3]`. This demand for an object is translated by calling `GEE(variable, context)` in the constructor in the translated Java class. This mechanism is in fact defining *demand-driven constructors*.

## 4.4 Inheritance

In the JOOIP semantics presented in Figure 1, we only allow embedded Lucid expressions to refer to members defined locally in the current class. We chose to do so out of simplicity of the expression of the semantics and its implemented solution that parses the JOOIP classes and extracts its local members in order to process the first semantic checking/translation step. Java reflection provides a powerful mechanism for the extraction of class members, even across an inheritance tree. That can then be used to do semantic checking taking into consideration an inheritance tree. However, reflection can only be used on standard Java classes, which JOOIP classes are not. We could further extend the possibility of the use of inheritance in JOOIP classes by using reflexion in the semantic checking/translation steps occurring after the standard Java code has been generated.

## 4.5 Intensional Relationship Across Object of an Intensional Class

A standard Java object is an instance of a Java class. These objects are isolated (i.e. not related to each other, except by the fact that they belong to the same class, i.e. a standard Java class declares a set of object whose individual evaluation is unrelated to the others. JOOIP allow to implicitly define intensional relationships across objects of a JOOIP intensional class. JOOIP inherits the essential feature of Lucid that permits it to use intensional operators to express relationship between across a multidimensional context space, in turn permitting to define streams of elements whose values depend on other elements.

In JOOIP, since we have objects as first class values, we can in fact create intensional relationships among Java objects of a JOOIP class. By declaring an intensional data member inside a JOOIP intensional class, we in fact implicitly define an intensional relationship *across* the objects of this class.

# 5 Examples of Application

Having described the syntax and semantics of JOOIP, then discussed some important issues of the resulting hybrid paradigm, we have kept our explanations at a conceptual level that maybe does not allow the reader to see concretely how can JOOIP be used to write programs, how such programs are effectively translated, and what are the advantages brought forth by the use of JOOIP. This section presents simple JOOIP program examples that will concretely demonstrate the capacities of JOOIP, as well as illustrate its translation process and its resulting output code. Note that three other examples are presented in Section 6 as a means to illustrate the advantages of JOOIP over GLU# [16].

## 5.1 Simple Example Illustrating the JOOIP-to-Java Translation Process

Listing 1 shows the typical natural numbers example derived from [17, 12] of Lucid written in JOOIP. The problem is to extract a value from the stream representing the natural numbers, beginning from the ubiquitous number 42. Let us arbitrarily pick the third value of the stream and set the stream's variance in the $d$ dimension; then in the `main()` method, the tag number two will be assigned to the method `computerLocalAverage()`. The method invocation `computerLocalAverage(2)` will ask the average value of its neighbors, which is the average of the second and forth value of the stream. With not much intuition, one can readily expect the program to return the value 44 which is the result of calculation $(43+45)/2$. The program will start with the Java `main()` method which will call `computeLocalAverage(int f)` method. According to the semantic translation process, the Lucid expression embedded in `computeLocalAverage(int f)` changes to a method calls, `GEE.eval([d:f-1],N)`, the expression will become: `GEE.eval([d:f+1],N)`. Here we will pass parameter `f` as context for Lucid expression. The run-time engine will drive two demands to calculate the value of the first and third element of the stream `N`. The definition of `N` is already stored in the corresponding GEER. According to the definition of `N`, after executing, the engine returns the `GIPLtest` object with value 44. The method `print()` will display the result on the screen. The translation process can be abstractly described as the following:

- According to the semantic rules, the JOOIP processor parses this program first and extracts all Lucid code segments.
- The JOOIP compiler then passes the three Lucid segments to the right GIPL parser to get the corresponding GEERs.
- Then, the JOOIP compiler passes the `GIPLtest.jop` program to the Java Parser to get the class's symbol table.
- With the class symbol table and GEERs in hand, the JOOIP compiler calls upon the GIPSY's Semantic Analyzer to do semantic check for each Lucid code segment.
- If the semantic check passes, at the last step, the JOOIP compiler generates a standard Java class, which is shown in Listing 2.

Notice here that, following our analysis/translation method, the embedded Lucid expressions ought to be "self-contained". As it is now, upon semantic check, the processor will report "undefined dimension

```
class GIPLtest
{
    private int N = @/#GIPL
                    if (#.d)<=0 then 1 else (N+1)@[d:(#.d-1)] fi
                    where
                      dimension d;
                    end /@

    public int computeLocalAverage(int f)
    {
        return  ( /@ N@.d f - 1 where dimension d; end @/
                + /@ N@.d f + 1 where dimension d; end @/) / 2;
    }

    public void print()
    {
        System.out.println("N=" + N);
    }

    public static void main(String[] argv)
    {
        GIPLtest oTest = new GIPLtest();
        oTest.N = oTest.computeLocalAverage(2);
        oTest.print();

    }
}
```

Listing 1: Natural numbers example written in JOOIP

d" for expressions using a dimension without a corresponding dimension declaration inside the `where` clause. If we add the `where` clause that includes the dimension declaration for each Lucid expression, after translation, the pure Java program will be like shown in Listing 2. That can be fixed by allowing our semantic checker to have awareness of the dimensionality of the intensional data members defined in the class, thus eliminating the need to redeclare dimensions in each embedded Lucid expression that refers to these dimensions. Once the JOOIP class has been translated to a regular Java class, the standard Java compiler will take care of the program compilation and running in within the JVM will print the result on the screen. Since the program also implements the `ISequentialThread` interface, it can be run by the GEE as well as being transported as demands distributedly. The above `GIPLtest.java` program is already correctly parsed by the Java parser and is smoothly integrated with GEE. From output, it shows that the `GIPLtest.java` calls engine and generate correct ASTs and dictionary.

## 5.2 Euler and Feynman Algorithms in JOOIP

The Euler and Feynman algorithms are very famous in physics. This section will show how to use JOOIP to apply the Euler and Feynman algorithm as well as the comparison with a traditional implementation in Java. By Newton's second law [15] we can calculate the acceleration of a body once we know the forces acting on it. The forces are either contact forces or field forces and may vary with time, position and velocity as the body moves. In order to describe the kinematics of the motion we need expressions for where the particle is and how fast it is moving at any time. A table which consists of values of position and velocity at specified time intervals is a numerical approach to kinematics. The accuracy of the tabled values depends on the approximations involved in their calculation. In the Euler algorithm, the average velocity and acceleration are replaced by the velocity and acceleration at the beginning of the interval as the equations (1) and (2) where $t$ is the beginning of the time interval, $dt$ is time interval, $v$ is velocity, $a$ is acceleration and $x$ is position:

$$v(t+dt) = v(t) + a(t) \cdot dt \quad (1)$$
$$x(t+dt) = x(t) + v(t) \cdot dt \quad (2)$$

The values at the beginning of the interval are known, and although they are not the best approxima-tion for the average values, they are not bad if the time interval is short enough. The Feynman algorithm approximates the average acceleration and velocity over a time interval by their values at the midpoint (in time). The equations on which the Feynman algorithm are based can be written as equations (3) and (4) using the same notations as for equations (1) and (2).

$$x(t+dt) = x(t) + v\left(t+\frac{dt}{2}\right)\cdot dt \tag{3}$$

$$v\left(t+\frac{dt}{2}\right) = v\left(t-\frac{dt}{2}\right) + a(t)\cdot dt \tag{4}$$

In equations (3) and (4), changes in position are calculated using a velocity value that is half a step ahead in time. Likewise, changes in velocity are calculated using an acceleration which is half a step ahead in time. Position and acceleration are therefore "in-phase" that is, they are calculated at the same points in time, and velocity is stepped half a step out of phase with both position and acceleration. We can use the Euler and Feynman algorithms to follow the motion of a mass on a spring with assumption that acceleration depends only on time and position. Consider a mass of $2kg$ attached to a spring with a force-constant of $8N/m$. It is passing through its equilibrium position at a velocity of $2.8m/s$ at time zero, and we want to follow its motion for one second with a time step of $0.2 seconds$. In [15], a Java program is provided. The main part is shown in Listing 3. We can find the above Java code has the following limitations:

1. the expression is very confusing, it cannot convey the natural meaning of the original differential equations easily;
2. infinity cannot be expressed in Java code, we have to pose a beginning and end of time explicitly;
3. it is not very good at describing of kinematics of the motion, because we cannot arbitrarily ask for information at a time, i.e. it has a purely extensional view and model of computation;
4. if forces depend on velocity, at the time acceleration is calculated, the only value available for velocity is the one from a half-step earlier. In this algorithm, the acceleration cannot depend on velocity;

If we assume that time interval $T$, position $Y$, acceleration $A$ and velocity $V$ are streams, we find they have same dimension *time* and the process is shown in Figure 4. From Figure 4, we find that the Euler is easier to program in Lucid and results in a program that resembles much more the original differential equations. However, there is more trouble with the Feynman algorithm because the data is "out-of-phase". If we consider each group of data as an individual object on the same dimension, objects have different structure. This also conflicts with the semantic restriction that each element of a stream has same type. Object-oriented concept in the hybrid JOOIP language helps to resolve the trouble. What needs to be done is to encapsulate same structure objects into a class, and then we will have two classes, class `InPhase` and class `OutPhase`. Even, we can let objects from the two classes to have the same context but different values. Figure 5 shows the solution.

Even if we keep track of changes in velocity as well as velocity itself and using the last change to project forward a half-step for the acceleration calculation [15], we can include another $V$ stream in class $A$ to record the velocity at the same time as $Y$ and $A$. This also fixes the "out-of-phase" problem. The program is in Listing 4.

In Figure 5 the reason why we need two classes, `InPhase` and `OutPhase`, it is because the distance `Y`, acceleration `A` and velocity `V` do not change at the same time. The two former change at `[time:0.2]`, while the latter changes at `[time:0.1]`. Moreover, acceleration `A` and velocity `V` depend on each other even though they don't change with the same pace. Let us go back to check the program: identifiers `T`, `Y` and `A` in class `InPhase` are three intensional streams which stand for time interval, distance and acceleration; identifiers `T` and `V` in class `OutPhase` are two intensional streams which stand for time interval and velocity. The dependencies among these identifiers are shown in above figure, for context `[time:t]`, distance `Y` in `InPhase` depends on the velocity `V` in `OutPhase` at the same context; acceleration `A` in `InPhase` depends on distance `Y` in `InPhase` at the same context; velocity `V` in `OutPhase` depends on the acceleration `A` in `InPhase` at context `[time: t-1]`. The program starts with the `main()` method. The run-time engine will generate three demands in class `InPhase`, they are `eval([time:3],T)`, `eval([time:3],Y)` and `eval([time:3],A)` and one demand in class `OutPhase` which is `eval([time:3],V)`. The definition of `T`, `Y`, `A` and `V` are already stored in the corresponding GEER. By object access and intensional operator, the program can easily get correct values while not like traditional sequential programs. The method `output()` will display the result on the screen. We find the following advantages by using JOOIP for this program:

1. A stream could be infinite, and is in fact defining the *intension* of the series of values, as opposed to the Java program that computes the *extension* of a portion of the infinite stream. With the intensional description available, we can ask kinematics of the motion at any time or interval of time in this intensional definition, yielding an extensional finite portion of the infinite stream;

2. It fixes the "out-of-phase" problem and can be used in the velocity-dependent case;

3. The potential parallelism in the computation is exploited by the eductive model of computation as implemented in the GEE and GIPSY;

4. Intensional objects are organized as a group;
5. Redundant computation is avoided by an intensional value warehouse;
6. The same intensional description can be used to compute values in any context by changing the initial demand;
7. It provides Lucid with richer input/output capabilities provided by the Java counterpart;

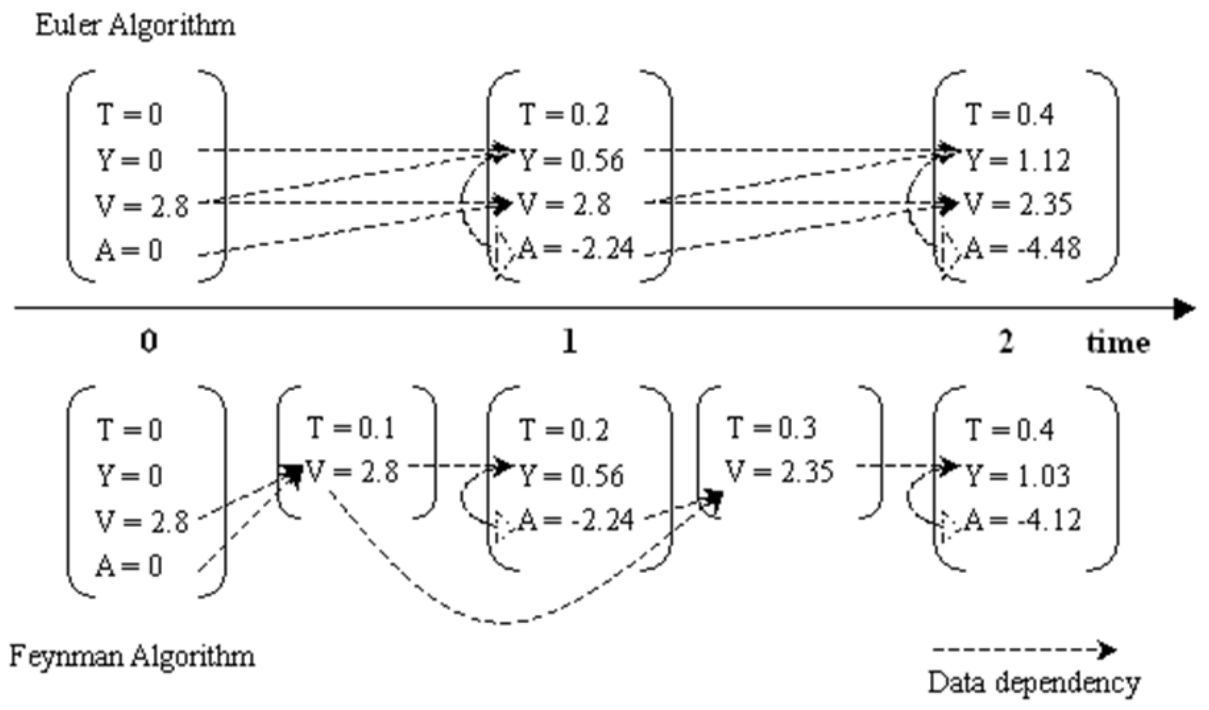

Figure 4: The process of generating data by both algorithms

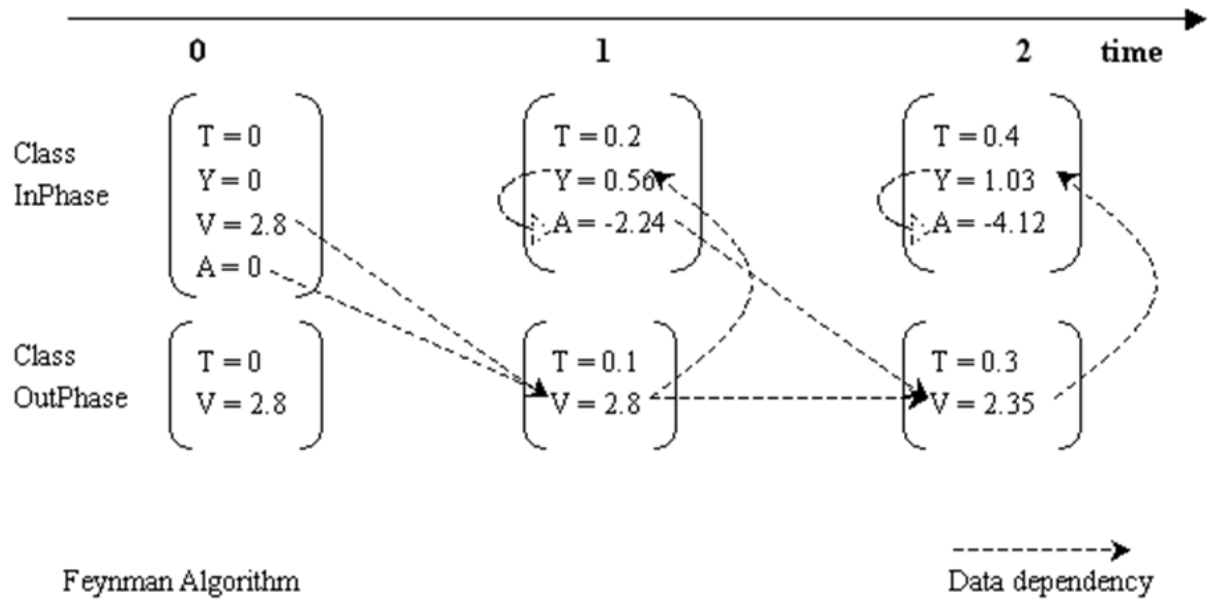

Figure 5: Solution for the Feynman algorithm with JOOIP

# 6 Related Work

The two paradigms used in this work have a generally poor interface among each other: on the one hand are conventional imperative programming languages that have no room for multidimensionality or intensional or demand-driven evaluation; on the other hand, existing multidimensional languages that cannot take advantage of imperative features and techniques. Developed over years of research, the combination typically result in much better performance. The following solutions are typical combination in this domain.

### GLU# – Intensional Language and C++

This approach embeds a small multidimensional core (GLU#) in a mainstream object-oriented programming language (C++) [16]. By this way, without changing the syntax and semantics of the host language, multidimensionality can be supported. GLU# is a small subset of GLU and it is implemented as a collection of C++ classes and class templates. All operators in Lucid appear as C++ functions. GLU# does not support Lucid functions; however, programmers are able to use lazy method templates in C++ to use C++'s functions in GLU#. GLU# provides a bridge between Lucid and OO. However, it just includes the basic features of Lucid and embeds into C++. Moreover, in contrast to C++, which supports mutable variables and objects, multidimensional objects in GLU# are immutable.

To show the similarities and differences between our work on JOOIP and GLU#, we provide the translation of some of the example given in [16] into JOOIP for the comparison reasons and to show how our approach is more general, adaptive, and flexible than that of GLU#. The examples include the embedding of the Sieve of Eratosthenes (see Listing 5) and the Hamming numbers (see Listing 6) examples with Indexical Lucid as an intensional dialect as well as the `TrafficLight` example with embedded Objective Lucid dialect segment (see Listing 7).

For the `Prime` and `Hamming` examples, it is very easy for Lucid programmers to integrate Lucid code in a Java class in JOOIP, because Lucid code keep the same nature. You do not need to change Lucid expressions into function format which is the case in GLU#. For the `TrafficLight` example, the JOOIP code is a lot shorter and simpler than GLU#'s. Because of the flexibility of our existing framework, it is very easy for us to extend Lucid code to any Lucid dialect. This is not the case in GLU#, functions are not allowed in GLU#, any time there is need for functions, programmers have to add C++'s templates. The most important, `OBJECTIVELUCID` makes it is possible for Lucid code to access identifiers and methods of Java class by introducing the object-oriented dot operator, this also help JOOIP to become a 2-way hybrid intensional-imperative programming language. Which is also not the case in GLU#.

### Embedding Lucid as Lazy Multidimensional Arrays

A similar embedding of multidimensional characteristics in a conventional programming language has been proposed by Rondogiannis [22]. In his approach, Java is used as the host language and intensional languages are embedded into Java as a form of definitional lazy multidimensional arrays. This approach tries to introduce multidimensional lazy array into Java and let the two different paradigms benefit each other. However, according to [22], there are several limitations. We will provide JOOIP code from the same multi program (see Listing 8) to show the difference.

1. The embedded language supports only a subset of the dimensional operators of GLU, the user-defined functions and the nested `where` clauses are not allowed in the language.
2. The value of a specific element of a lazy array can not be altered by a procedural program. In JOOIP program, we pass `f` as parameter to be the tag of context `[x:f][y:f]` of stream `P`. This parameter pass as tag solution is already implemented in Natural number example.
3. Only one multidimensional array definition section is allowed in every class definition. In JOOIP, we don't limit the number of places where the IPL could appear.

However, this approach concentrates more on implementation efficiency improvement with lazy array evaluation comparing to imperative array instead of language design itself like JOOIP.

### JLucid

JLucid [12, 6] was a first attempt on intensional arrays and "free Java functions" in the GIPSY. The approach used the Lucid language as the driving main computation, where Java methods were peripheral and could be invoked from the Lucid part, but not the other way around. This was the first instance of hybrid programming within the GIPSY. The semantics of this approach was not completely defined, plus, it was only one-sided view (Lucid-to-Java) of the problem. JLucid did not support objects of any kind, but introduced the wrapper class idea for the free Java methods.

### Objective Lucid

Objective Lucid [12, 11] was a logical extension of the JLucid language mentioned in the previous section that inherited all of the JLucid's features and introduced Java objects to be available for use by Lucid. Objective Lucid expanded the notion of the Java object (a collection of members of different types) to the array (a collection of members of the same type) and first introduced the dot-notation in the syntax and operational semantics in GIPSY. Like in JLucid, Objective Lucid's focus was on the Lucid part being the "main" program and did not allow Java to call intensional functions or use intensional constructs from within a Java class. Objective Lucid was the first in GIPSY to introduce the more complete operational semantics of the hybrid OO intensional language. This work greatly complements Objective Lucid by allowing Java to call the intensional language constructs closing the gap and making JOOIP a complete hybrid OO intensional programming language within the GIPSY environment. JOOIP's semantics further refines in a greater detail the operational semantics rules of Lucid and Objective Lucid in the attempt to make them complete.

### GICF

The General Imperative Compiler Framework [10] is the GIPSY's compiler framework that allows for a generalized way of inclusion of any imperative languages into intensional variants within the GIPSY environment and allowing the syntactical co-existence of the intensional and imperative languages in one source file by providing a `Preprocessor` that splits the intensional and imperative code chunks that are fed to their respective compilers, and then the results are gathered and linked together to form a compiled hybrid program as an instance of GEER. This work benefits from some of the existing infrastructure put in place in the GIPSY to implement the compilation process.

### Object-Oriented Intensional Implementation

In [4], there is another discussion on issues about object-oriented implementation of intensional languages. In this approach, each variable in a Lucid program is considered as a class and an object of a class is a variable in a context. Each variable definition in a Lucid program is compiled into a C++ class definition that has the same name as the variable. This approach focuses on the implementation-level by creating a class for each Lucid variable, it helps the system to execute in a distributed manner. However, the objects introduced here does not contain information from C++ variables, which is provided by our proposed solution.

### Other Related Propositions

The concept about objects in Lucid first appeared in [5] in the early 1990s; however, it did not clearly define how to realize this idea. In the later 1990s, Peter Kropf and John Plaice talked about this topic in their paper "intensional objects" [9]. In this paper, intensional objects are considered as openable boxes labeled by Lucid contexts. This paper focuses on intensional versioning whose task is to build a system from versioned components, which are already sitting in the warehouse. This warehouse is different as the warehouse in intensional programming. The latter is like a cache to improve the performance. The former contains the source of everything, it is like a "catalog" or a "repository", in which the boxes are put. Each box is of some contents and a tag that is context. Thus, in this approach, these labeled boxes are called intensional objects, which are re-openable and re-packageable. However, in this approach, the authors did not clearly define the relationship among these "boxes" and if these boxes could include intensional concepts. Moreover, the idea is only on conceptual level.

## 7 Conclusion

The translation process will limit certain qualities of our solution, e.g. the resulting implementation would inevitably be less efficient than if directly translated into object code. We have to keep in mind, however, that this work's main goal is language development and a proof-of-concept of the developed language. Execution optimization will come in time if our language concepts prove to be fruitful.

By integrating Lucid and Java, we combine the essential characteristics of object-oriented languages with the basic elements of Lucid. We make it possible that each element in a stream could be an object. By this new point, we extend the use of objects and enrich the meaning of a stream in Lucid, which can greatly increase the power of Lucid. The hybrid OO-IP approach proposed here and adopted by JOOIP is enabling the novel concept of *intensional member* and *intensional class*.

The use of the GICF compiler framework has proven to meet its flexibility promises in this work by greatly facilitating the development of a hybrid language permitting the use of any variant of Lucid to be embedded in Java classes. By the use of this facility, JOOIP is the first language to allow the embedding of different variants of Lucid. Moreover, as soon as a new compiler for a variant of Lucid is added to the GICF, it is automatically available to JOOIP. This feat proves the validity of the framework approach adopted for the design of the GIPC.

As we were working on the design of JOOIP, it initially appeared that we would have to apply profound changes to our run-time system (GEE). Interestingly, the solution we present here did not require any such change, again proving the generality of our design. Generality and flexibility has been one of our major goals in the design of the GIPSY. This research experience and result provides us with a positive evaluation of these requirements.

## 8 Acknowledgments

This work was funded by NSERC and the Faculty of Computer Science and Engineering, Concordia University. Thanks to many of the GIPSY project team members for their valuable contributions, suggestions, and reviews, including Peter Grogono, Emil Vassev, Xin Tong, and Amir Pourteymour.

```
package gipsy.tests.jooip;
import gipsy.GIPC.GIPC;
import gipsy.lang.*;
import gipsy.lang.converters.type.*;
import gipsy.interfaces.*;
import gipsy.lang.context.Dimension;
import gipsy.GEE.GEE;
import gipsy.util.*;
import java.lang.reflect.Method;

public class GIPLtest implements ISequentialThread
{
        private static GIPSYProgram soGEER1;
        private static GIPSYProgram soGEER2;
        private static GIPSYProgram soGEER3;
      private int N = 0;
        private boolean bNIsWritten = false;
        private GIPSYContext oContext;

        public GEE oGEE3 = new GEE(soGEER3);
        public GEE oGEE2 = new GEE(soGEER2);
        public GEE oGEE1 = new GEE(soGEER1);

        static{
                try
                {
                        GIPC oGIPC1 = new GIPC(new StringInputStream("N  where N = if (#.d) <= 0 then 1 else (N+1) @.d (#.d) −
                               1 fi                              where  dimension d;   end ; end"), new String[] {"−−gipl", "−−debug"
                              });
                        oGIPC1.compile();
                        soGEER1 = oGIPC1.getGEER();
                        GIPC oGIPC2 = new GIPC(new StringInputStream(" N@.d f − 1 where dimension d; end "), new String[] {"−−
                              gipl", "−−debug"});
                        oGIPC2.compile();
                        soGEER2 = oGIPC2.getGEER();
                        GIPC oGIPC3 = new GIPC(new StringInputStream(" N@.d f + 1 where dimension d; end "), new String[] {"−−
                              gipl", "−−debug"});
                        oGIPC3.compile();
                        soGEER3 = oGIPC3.getGEER();
                }
                catch(Exception e)
                {
                        System.err.println(e);
                        e.printStackTrace(System.err);
                }
        }

        public GIPLtest(GIPSYContext poContext)
        {
                this.oContext = poContext;
        }

        public GIPLtest()
        {
        }

        public WorkResult work()
        {
                GIPLtest.main(null);
                return null;
        }
        @Override
        public WorkResult getWorkResult()
        {
                return null;
        }

        @Override
        public void setMethod(Method poSTMethod)
        {
        }

        @Override
        public void run()
        {
                work();
        }
/**
 * @version $Id: jooip−09.tex,v 1.8 2009/09/03 05:43:20 mokhov Exp $
 */


    public int computeLocalAverage(int f)
    {
                this.oContext = new GIPSYContext();
                Dimension oDimension = new Dimension();
                oDimension.setDimensionName(new GIPSYIdentifier("d"));
                oDimension.setCurrentTag(new GIPSYInteger(f));
                this.oContext.addDimension(oDimension);

       return  ( IPLToJava.convertToInteger(this.oGEE2.eval(this.oContext))                  + IPLToJava.convertToInteger(this.
            oGEE3.eval(this.oContext))) / 2;
    }

    public void print()
    {
        System.out.println("N=" + N);
    }

    public static void main(String[] argv)
    {
        GIPLtest oTest = new GIPLtest();
```

```
k=8; m=2; dt=0.2;
t=0; y=0; v=2.8; a=-k*y/m;    // Euler Algorithm
for (int i=1; i<=5; i++) {
  t=t+dt;
  y=y+v*dt;
  v=v+a*dt;
  a=-k*y/m;
}

t=0; y=0; v=2.8; a=-k*y/m;    // Feynman Algorithm
v=v+a*dt/2;
for (int i=1; i<=5; i++) {
  t=t+dt;
  y=y+v*dt;
  a=-k*y/m;
  v=v+a*dt;
}
```

Listing 3: Java code for the Euler and Feynman Algorithms

```
public class InPhase
{
  InPhase  C1 = new InPhase();
  OutPhase C2 = new OutPhase();

  int k = 8;
  int m = 2;

  double T = /@#INDEXICALLUCID 0 fby.time (T + 0.2)
                  Where dimension time; end @/;
  double Y = /@#OBJECTIVELUCID 0 fby.time (Y + C2.V @.time #.time * 0.2) where dimension time; end @/;
  double A = /@#OBJECTIVELUCID 0 fby.time (-k / m) * Y @.time #.time where dimension time; end @/;

  public void output(double interval, double distance, double speed, double accel)
  {
    System.out.println("Feynman Algorithm");
    System.out.println("Time = " + interval);
    System.out.println("Position = " + distance);
    System.out.println("Acceleration = " + accel);
    System.out.println("Velocity = " + speed);
  }

  public static void main(String[] argv)
  {
    InPhase oMotionIn = new InPhase();
    OutPhase oMotionOut = new OutPhase();

    double T_value = /@#OBJECTIVELUCID oMotionIn.T @.time 3
                            where dimension time; end @/;
    double Y_value = /@#OBJECTIVELUCID oMotionIn.Y @.time 3
                            where dimension time; end @/;
    double A_value = /@#OBJECTIVELUCID oMotionIn.A @.time 3
                            where dimension time; end @/;
    double V_value = /@#OBJECTIVELUCID oMotionOut.V @.time 3
                            where dimension time; end @/;
    oMotion.output(T_value, Y_value, A_value, V_value);
  }
}

public class OutPhase
{
  InPhase C3 = new InPhase();

  double T = /@#INDEXICALLUCID if (#.time==0) then 0 else (-0.1 fby.time (T+0.2)) fi  where dimension time; end @/;

  double V = /@#OBJECTIVELUCID if ((#.time == 0) || (#.time == 1)) then 2.8 else V + (C3.A @.time #.time - 1) * 0.2 fi
   where dimension time; end @/;
```

Listing 4: The program in JOOIP for Feynman Algorithm

```
class Prime
{
    private int prime = /@#INDEXICALLUCID
                          first.x sieve
                          where
                              dimension x, y;
                              sieve = ints fby.y (sieve wvr.x sieve % prime != 0);
                              where
                                  ints = 2 fby.x ints + 1;
                              end
                          end@/;

    int num;

    public void print()
    {
        System.out.println(prime);
    }

    public static void main(String[] argv)
    {
        for (num=0; num<100; num++) /@ prime@[x:num] @/.print();
    }
}
```

Listing 5: The `Prime` example rewritten in JOOIP

```
class Hamming
{
    private int H = /@#INDEXICALLUCID
                        1 fby.d merge.d(merge.d(2*H, 3*H), 5*H);
                        where
                            dimension d;
                            merge.a (x,y) = if (xx<=yy) then xx else yy fi
                                where
                                    dimension a;
                                    xx = x upon.a (xx<=yy);
                                    yy = y upon.a (yy<=xx);
                                end
                        end@/;

    int num;

    public void print()
    {
        System.out.println(H);
    }

    public static void main(String[] argv)
    {
        for (num=0; num<100; num++) /@ H@[d:num] @/.print();
    }
}
```

Listing 6: The `Hamming` example rewritten in JOOIP

```
class TrafficLight
{
    String[] StateName = new String[]{"GREEN", "YELLOW", "RED"};
    int[] timePerLight = new int[]{5, 1, 8};

    int getState(String statename)
    {
        if (statename.equals("GREEN")) return 0;
        else if (statename.equals("YELLOW")) return 1;
        else if (statename.equals("RED")) return 2;
    }

    public String state = "RED";
    public int timer = 8;

    TrafficLight statu = new TrafficLight();

    private String light = /@#OBJECTIVELUCID
                                statu.state fby.t statu.next();
                                where
                                    dimension t;
                                end@/;

    int num;

    public TrafficLight()
    {
        this.state = state;
        this.timer = timer;
    }

    String next()
    {
        int t = timer - 1;
        int position;
        String LightColor = "";

        if(t <= 0)
        {
            position = (getState(state)+1) % 3;
            t = timePerLight[position];
            state = StateName[position];
        }

        timer = t;
        LightColor = state;

        return LightColor;
    }

    public void print()
    {
        System.out.println(light);
    }

    public static void main(String[] argv)
    {
        for (num=0; num<100; num++) /@ light@[t:num] @/.print();
    }
}
```

Listing 7: The `TrafficLight` example rewritten in JOOIP

```
public class multi
{
    private int P = /@#INDEXICALLUCID
                        1 fby.x (0 fby.y (next.y(p)+next.x(p))/2)
                    where
                        dimension x,y;
                    end @/;

    public void print(f)
    {
        System.out.println(/@LUCX P @[x:f][y:f] @/);
    }

    public static void main(String[] argv)
    {
        GIPLtest oTest = new multi();
        oTest.print(2);
    }
}
```

Listing 8: The `multi` example rewritten in JOOIP